\documentclass[pra,showpacs,showkeys]{revtex4}

\usepackage{graphicx}
\usepackage{dcolumn}
\usepackage{bm}

\begin{document}

\title{3D+1 Lorentz type soliton in air}
\author{Lubomir M. Kovachev}
\affiliation{Institute of Electronics, Bulgarian Academy of
Sciences, Tzarigradcko shossee 72,1784 Sofia, Bulgaria,
e-mail:lubomirkovach@yahoo.com}

\date{\today}

\begin{abstract}
Up to now the long range filaments have been considered as a balance
between Kerr focusing and defocusing by plasma generation in the
nonlinear focus.  However, it is difficult to apply the above
explanation of filamentation in far-field zone. There are basically
two main characteristics which remain the same at these distances -
the super broad spectrum and the width of the core, while the power
in a stable filament drops to the critical value for self-focusing.
At such power the plasma and higher-order Kerr terms are too small
to prevent self-focusing. We suggest here a new mechanism for stable
soliton pulse propagation in far-away zone, where the power of the
laser pulse is slightly above the critical one, and the pulse
comprises super-broad spectra. For such pulses the diffraction is
not paraxial and an initially symmetric Gaussian pulse takes
parabolic form at several diffraction lengths . The stable soliton
propagation appears as a balance between the divergent parabolic
type diffraction of broadband optical pulses and the convergent
nonlinear refractive index due to the intensity profile. We
investigate more precisely the nonlinear third order polarization,
using  into account the carrier-to envelope phase. This additional
phase transforms the third harmonic term to THz or GHz one,
depending on the spectral width of the pulse.
\end{abstract}

 \maketitle

\section{Introduction}
In the process of investigating the filamentation of a power
femtosecond (fs) laser pulse many new physical effects have been
observed, such as long-range self-channeling \cite{MOU, WO, MEC},
coherent and incoherent radial and forward THz emission \cite{TZ,
DAMYZ, DAHO}, asymmetric pulse shaping, super-broad spectra
\cite{HAU1, HAU2, COU1, SKUP, CHIN1, KASP1} and others. The role of
the different mechanisms in near zone (up to $1-2$ $m$ from the
source) has been investigated experimentally and by numerical
simulations, and most  processes in this zone are well explained
\cite{COU, KAN, FAC, KOL1}. When a fs pulse with power of several
$P_{cr}=\pi(0.61\lambda_0)^2/(8n_0n_2)$ starts from the laser source
a slice-by-slice self-focussing process takes place \cite{SHEN}. At
a distance of one-two meters the pulse self-compresses, enlarging
the $k_z$ spectrum to super-broad asymmetric spectrum $\triangle
k_z\approx k_0$. The process increases the core intensity up to tens
of $10^{13} W/cm^2$, where different types of plasma ionization,
multi-photon processes and higher-order Kerr terms appear
\cite{KASP2}. Usually, the basic model of propagation in near the
zone is a scalar spatio-temporal paraxial equation including all the
above mentioned mechanisms \cite{COU, KAN, KASP2}. The basic model
is natural in the near zone because of the fact that the initial fs
pulse contains a narrow-band spectrum $\triangle k_z << k_0$. Thus,
the paraxial spatio-temporal model gives a good explanation of
nonlinear phenomena such as conical emission, X-waves, spectral
broadening to the high frequency region and others. In far-away zone
(propagation distance more than $2-3$ meters) plasma ionization and
higher-order Kerr terms are admitted also as necessary for a balance
between the self-focussing and plasma defocussing and for obtaining
long range self-channeling in gases.

However, the above explanation of filamentation is difficult to
apply in far-away zone. There are basically two main characteristics
which remain the same at these distances - the superbroad spectrum
and the width of the core, while the intensity in a stable filament
drops to a value of $10^{12} W/cm^2$ \cite{COU, KASP2}. The plasma
and higher-order Kerr terms are too small to prevent self-focussing.
The observation of long-range self-channeling \cite{MECH, DUB, BERN}
without ionization also leads to change the role of plasma in the
laser filamentation.

In addition, there are  difficulties with the physical
interpretation of the THz radiation as a result of plasma
generation. The plasma strings formed during filamentation should
emit incoherent THz radiation in a direction orthogonal to the
propagation axis. The nature of the THz emission, measured in
\cite{DAHO} is different. Instead of being emitted radially, it is
confined to a very narrow cone in the forward direction. The
contribution from ionization in far-away zone is negligible
\cite{KASP2} and this is the reason to look for other physical
mechanism which could cause THz or GHz radiations. Our analysis on
the third order nonlinear polarization of pulses with broadband
spectrum indicates that the nonlinear term in the corresponding
envelope equation oscillates with frequency proportional to the
group and phase velocity difference
$\Omega_{nl}=3(k_0v_{ph}-v_{gr}\triangle k_z)$. Actually, this is
three times the well-known Carrier-to Envelope Phase (CEP)
difference \cite{EXTR}. This oscillation induces THz generation,
where the generated frequency is exactly $\Omega_{THz}=93 GHz$ for a
pulse with superbroad spectrum $\triangle k_z\approx k_0$ with
carryier wavelength $800$ $nm$.

Physically, one dimensional Schr\"{o}dinger solitons in fibers
appear as a balance between the Kerr nonlinearity and the negative
dispersion \cite{HAS, AGR, SER1}. On the other hand, if we try to
find 2D+1 and spatio-temporal solitons in Kerr media, the numerical
and the real experiments demonstrate that there is no balance
between the plane wave paraxial diffraction - dispersion  and the
Kerr nonlinearity. This leads to instability and self-focusing of a
laser beam or initially narrow band optical pulse. Recently Serkin
in \cite{SER2} suggested stable soliton propagation and reducing the
3D soliton problem to one dimensional, with introducing trapping
potential in Bose - Einstein condensates.

In this paper we present a new mathematical model, on the basis of
the Amplitude Envelope (AE) equation, up to second order of
dispersion,  without using paraxial approximation. In the
non-paraxial zone the diffraction of pulses with superbroad spectrum
or pulses with a few cycles under the envelope is closer to wave
type \cite{KOV1}. For such pulses, a new physical mechanism of
balance between nonparaxial (wave-type diffraction) and third order
nonlinearity appears. Exact analytical three-dimensional bright
solitons in this regime are found.

\section{Linear regime of narrow band and broad band  optical pulses}

The paraxial spatio-temporal envelope equation governs well the
transverse  diffraction and the dispersion of fs pulses up to  $6-7$
cycles under the envelope. This equation relies on one approximation
obtained after neglecting the second derivative in the propagation
direction and the second derivative in time from the wave equation
\cite{CHRIS} or from the $3D+1$ AE equation \cite{BRAB}. In air, the
series of $k^2(\omega)$ are strongly convergent up to one cycle
under the envelope and this is the reason why the AE equation is
correct up to the single-cycle regime.

The linearized AE, governing the propagation of laser pulses when
the dispersion is limited to second order, is:

\begin{eqnarray}
\label{DDE}  -2ik_0\left(\frac{\partial A}{\partial z}+
\frac{1}{v_{gr}}\frac{\partial A}{\partial t}\right) = \Delta A
-\frac{1+\beta}{v_{gr}^2}\frac{\partial^2 A}{\partial t^2},
\end{eqnarray}
where $\beta=k"k_0v_{gr}^2 $ is a number representing the influence
of the second order dispersion. In vacuum and dispesionless media
the following Diffraction Equation (DE) ($v\sim c$ ) is obtained:

\begin{eqnarray}
\label{DE}  -2ik_0\left(\frac{\partial V}{\partial z}+
\frac{1}{v}\frac{\partial V}{\partial t}\right) = \Delta V
-\frac{1}{v^2}\frac{\partial^2 V}{\partial t^2}.
\end{eqnarray}
We solve AE (\ref{DDE}) and DE (\ref{DE}) by applying spatial
Fourier transformation to the amplitude functions $A$ and $V$. The
fundamental solutions of the Fourier images $\hat{A}$ and $\hat{V}$
in ($k_x,k_y,\triangle k_z ,t$) space are:

\begin{eqnarray}
\label{DDK} \hat{A}=\hat{A}(k_x,k_y,\triangle k_z ,t=0)\times\nonumber\\
\exp\left\{i\frac{v_{gr}}{\beta+1}\left(k_0
\pm\sqrt{k_0^2+(\beta+1)\left({k_x}^2+{k_y}^2+{\triangle k_z
}^2-2k_0 \triangle k_z \right)}\right)t\right\},
\end{eqnarray}

\begin{eqnarray}
\label{DK} \hat{V}=\hat{V}(k_x,k_y,\triangle k_z ,t=0)
\exp\left\{iv\left(k_0 \pm\sqrt{{k_x}^2+{k_y}^2+({\triangle k_z
}-k_0 )^2}\right)t\right\},
\end{eqnarray}
respectively. In air $\beta\simeq 2.1\times10^{-5}$, AE (\ref{DDE})
is equal to DE (\ref{DE}), and the dispersion is negligible compared
to the diffraction. We solve analytically the convolution problem
(\ref{DK}) for initial Gaussian light bullet of the kind
$V(x,y,z,t=0)=\exp\left(-(x^2+y^2+z^2)/2r_0^2\right)$. The
corresponding solution is:

\begin{eqnarray} \label{EXACT1}
V(x,y,z,t)=\frac{i}{2\hat{r}}\exp\left[-\frac{k_0^2r^2_0}{2}+ik_0(vt-z)\right]\times\nonumber\\
\Biggr\{i(vt+\hat{r})\exp\left[-\frac{1}{2r^2_0}(vt+\hat{r})^2\right]
erfc\left[\frac{i}{\sqrt{2}r_0}(vt+\hat{r})\right]\\
-i(vt-\hat{r})\exp\left[-\frac{1}{2r^2_0}(vt-\hat{r})^2\right]
erfc\left[\frac{i}{\sqrt{2}r_0}(vt-\hat{r})\right]\Biggr\}\nonumber,
\end{eqnarray}
where $\hat{r}=\sqrt{x^2+y^2+(z-ir^2_0k_0)^2}$.  On the other hand,
multiplying the solution (\ref{EXACT1}) with the carrier phase, we
obtain solution of the wave equation $ E\left(x,y,z,t\right)=
V\left(x,y,z,t\right)\exp{\left(i(k_0z-\omega_0t)\right)}$, where
$\omega_0$ and $k_0 $  are the carrier frequency and carrier wave
number in the wave packet:

\begin{eqnarray}
\label {W} \Delta E= \frac{1}{v^2}\frac{\partial^2 E}{\partial t^2},
\end{eqnarray}

\begin{eqnarray} \label{EXACT2}
E(x,y,z,t)=\frac{i}{2\hat{r}}\exp\left(-\frac{k_0^2r^2_0}{2}\right)\times\nonumber\\
\Biggr\{i(vt+\hat{r})\exp\left[-\frac{1}{2r^2_0}(vt+\hat{r})^2\right]
erfc\left[\frac{i}{\sqrt{2}r_0}(vt+\hat{r})\right]\\
-i(vt-\hat{r})\exp\left[-\frac{1}{2r^2_0}(vt-\hat{r})^2\right]
erfc\left[\frac{i}{\sqrt{2}r_0}(vt-\hat{r})\right]\Biggr\}\nonumber.
\end{eqnarray}
A systematic study on the different kinds of exact solutions and
methods for solving  wave equation (\ref{W}) was performed recently
in \cite{KIS}. Here, as in \cite{KOV1} we suggest another method:
Starting with the ansatz $E\left(x,y,z,t\right)=
V\left(x,y,z,t\right)\exp{\left(i(k_0z-\omega_0t)\right)}$, we
separate the main phase and reduce the wave equation to $3D+1$
parabolic type one (\ref{DE}). Thus, the initial value problem can
be solved and exact (\ref{EXACT1}) (or numerical) solutions of the
corresponding amplitude equation (\ref{DE}) can be obtained. The
solution (\ref{EXACT1}), multiplied by the main phase, gives an
exact solution (\ref{EXACT2}) of the wave equation (\ref{W}). To
investigate the evolution of optical pulses at long distances, it is
convenient to rewrite AE (\ref{DDE}) equation in Galilean coordinate
system $t' = t; z' = z - v_{gr}t$:

\begin{eqnarray}
\label{gal} -i\frac{2k_0}{v_{gr}}\frac{\partial A}{\partial t'}=
\Delta_{\bot} A - \beta\frac{\partial^2 A}{\partial z'^2}
-\frac{1+\beta}{v_{gr}^2}\left(\frac{\partial^2 A}{\partial
t'^2}-2v_{gr}\frac{\partial^2 A}{\partial t'\partial z'}\right).
\end{eqnarray}
Pulses governed by DE (\ref{DE}) move with phase velocity and the
transformation is $t' = t; z' = z - vt$:

\begin{eqnarray}
\label{galvac} -i\frac{2k_0}{v}\frac{\partial V}{\partial t'}=
\Delta_{\bot} V-\frac{1}{v^2}\left(\frac{\partial^2 V}{\partial
t'^2}-2v\frac{\partial^2 V}{\partial t'\partial z'}\right).
\end{eqnarray}
Here, $\Delta_{\bot}=\frac{\partial^2}{\partial x^2} +
\frac{\partial ^2}{\partial y^2}$ denotes the transverse Laplace
operator. The corresponding fundamental solution  of AE equation
(\ref{gal}) in Galilean coordinates  is:

\begin{eqnarray}
\label{SKB}\hat{A}_G(k_x,k_y,\triangle k_z ,t)=\hat{A}_G(k_x,k_y,\triangle k_z ,t=0)\times\nonumber\\
\\
\exp\Biggr\{i\frac{v_{gr}}{\beta+1}\left[ k_0-(\beta+1)\triangle k_z
\pm\sqrt{\left(k_0-(\beta+1)\triangle k_z
\right)^2+(\beta+1)(k_x^2+k_y^2- \beta \triangle k_z
^2)}\right]t\Biggr\}\nonumber,
\end{eqnarray}
while the fundamental solution of DE (\ref{galvac}) becomes:

\begin{eqnarray}
\label{SDE}\hat{V}_G=\hat{V}_G(k_x,k_y,\triangle k_z ,t=0)\times\nonumber\\
\\
\exp\Biggr\{iv\left[ k_0-\triangle k_z  \pm\sqrt{\left(k_0-\triangle
k_z \right)^2+k_x^2+k_y^2}\right]t\Biggr\}\nonumber.
\end{eqnarray}
The analytical solution of (\ref{SDE}) for initial pulse in the form
of Gaussian bullet is the same as (\ref{EXACT1}), but with new
 radial component
$\hat{r}=\sqrt{x^2+y^2+(z+vt-ir^2_0k_0)^2}$ translated in space and
time. The numerical and analytical solutions of AE (\ref{DDE}) and
DE (\ref{DE}) are equal to the solutions of the equations AE
(\ref{gal}) and DE (\ref{galvac}) in Galilean coordinates with only
one difference: in Laboratory frame the solutions translate in
$z$-direction , while in Galilean frame the solutions stay in the
centrum of the coordinate system.

The basic theoretical studies governed laser pulse propagation have
been performed in so called "local time" coordinates $z=z$;
$\tau=t-z/v_{gr}$. In order to compare our investigation with these
results, we need to rewrite AE equation (\ref{DDE}) for the
amplitude function $A$ in the same coordinate system. Thus Eq.
(\ref{DDE}) becomes:

\begin{eqnarray}
\label{loct} -2ik_0\frac{\partial A}{\partial z}= \Delta_{\bot} A +
\frac{\partial^2 A}{\partial z^2} -\frac{2}{v_{gr}}\frac{\partial^2
A}{\partial \tau\partial z}-\frac{\beta}{v_{gr}^2}\frac{\partial^2
A}{\partial \tau^2}.
\end{eqnarray}
Since this is a parabolic type equation with low order derivative on
$z$, we apply Fourier transform to the amplitude function in form:
$\hat{A}\left(k_x,k_y,\triangle\omega,z\right)=FFF\left[A\left(x,y,z,t\right)\right]$,
where $FFF$ denotes 3D Fourier transform in $x,y,\tau$ space and
$\triangle\omega=\omega-\omega_0$; $\triangle
k_z=\triangle\omega/v_{gr}$ are the spectral widths in frequency and
wave vector domains correspondingly. The following ordinary
differential equation in $\left(k_x,k_y,\triangle\omega,z\right)$
space is obtained:

\begin{eqnarray}
\label{locF}
-2i\left(k_0-\frac{\triangle\omega}{v_{gr}}\right)\frac{\partial
\hat{A}}{\partial z}=
-\left(k_x^2+k_y^2-\frac{\beta\triangle\omega^2}{v_{gr}^2}\right)\hat{A}+\frac{\partial^2
\hat{A}}{\partial z^2}.
\end{eqnarray}
As can be seen from (\ref{locF}), if  the second derivative on $z$
is neglected,then the paraxial spatio-temporal approximation is
valid. Equation (\ref{locF}) is more general and we will estimate
where we can apply spatio-temporal paraxial optics (PO), and where
PO does not works.  The fundamental solution of (\ref{locF}) is:

\begin{eqnarray}
\label{locsol}
\hat{A}\left(k_x,k_y,\triangle\omega,z\right)=\hat{A}\left(k_x,k_y,\triangle\omega,0\right)\times\nonumber\\
\exp \left\{i\left[\left(k_0-\frac{\triangle\omega}{v_{gr}}\right)
\mp\sqrt{\left(k_0-\frac{\triangle\omega}{v_{gr}}\right)^2+k_x^2+k_y^2-
\frac{\beta\triangle\omega^2}{v_{gr}^2}}\right]z\right\}.
\end{eqnarray}
The analysis of the fundamental solution (\ref{locsol}) of the
equation (\ref{locF}) is performed in two basic cases:

a: Narrow band  pulses - from nanosecond up to $50-100$ femtosecond
laser pulses, where the conditions:
\begin{eqnarray}
\frac{\beta\triangle\omega^2}{v_{gr}^2}\leq k_x^2\sim k_y^2<<k_0^2;\
\triangle k_z=\frac{\triangle\omega}{v_{gr}}<<k_0
\end{eqnarray}
are satisfied, and the wave vector's difference
$k_0-\triangle\omega/v_{gr}$ can be replaced by $k_0$. Using the low
order of the Taylor expansion and the minus sign in front of the
square root from the initial conditions, equation (\ref{locsol}) is
transformed in a spatio - temporal paraxial generalization of the
kind:

\begin{eqnarray}
\label{fresnel}
\hat{A}\left(k_x,k_y,\triangle\omega,z\right)=\hat{A}\left(k_x,k_y,\triangle\omega,0\right)
\exp\left[i\left(\frac{k_x^2+k_y^2-\frac{\beta\triangle\omega^2}{v_{gr}^2}}{2k_0}\right)z\right].
\end{eqnarray}
From (\ref{fresnel}) the evolution of the narrow band pulses becomes
obvious: while the transverse projection of the pulses enlarges by
the Fresnel's law, the longitudinal temporal shape will be enlarged
in the same away, proportionally to the dispersion parameter
$\beta$. Such shaping  of pulses with initially narrow band spectrum
is demonstrated in Fig.1, where   the typical Fresnel diffraction of
the intensity profile (spot $(x,y)$ projection) is presented. The
numerical experiment is performed for  $100$ femtosecond Gaussian
initial pulse at $\lambda=800$ nm, $\Delta k_z<<k_0$, $z_0=30 \mu
m$, $ r_0(x,y)=60 \mu m$, with $37.5$ cycles under envelope
propagating in air ($\beta=2.1\times 10^{-5}$). The result is
obtained by solving numerically the inverse Fourier transform of the
fundamental solution (\ref{locsol}) of the AE equation in the local
time frame (\ref{loct}). The spot enlarges twice at one diffraction
length $z_{diff}=r_0^2k_0$. Fig. 2 presents the intensity side
$(x,\tau)$ projection of the same pulse. We should note that while
the spot ($(x,y)$ projection) enlarges considerably due to the
Fresnel law, the longitudinal time shape (the $\tau$ projection)
remains the same on several diffraction lengths from the small
dispersion in air. The diffraction - dispersion picture, presented
by the side $(x,\tau)$ projection of the pulse, gives idea of what
should happen in the nonlinear regime: the plane wave diffraction
with a combination of parabolic type nonlinear Kerr focusing always
leads to self-focusing for narrow-band ($\Delta k_z<<k_0$) pulses.
The same Taylor expansion for narrow band pulses can be performed to
fundamental solutions of the equation in Laboratory (\ref{DDK}) and
Galilean (\ref{SKB}) frames.

\begin{figure}
 \centering
\includegraphics[width=120 mm]{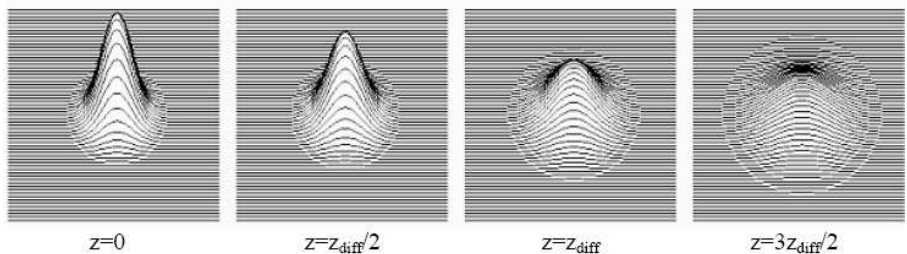}
\caption{Plot of the waist (intensity's) projection $|A(x,y)|^2$ of
a $100$ $fs$ Gaussian pulse at $\lambda=800$ $nm$, with initial spot
$r_0=60$ $\mu m$, and longitudinal spatial pulse duration $z_0=30$
$\mu m$, as solution of the linear equation in local time
(\ref{loct}) on distances expressed by diffraction lengths. The spot
deformation satisfies the Fresnel diffraction law and on one
diffraction length $z=z_{diff}$ the diameter of the spot increases
twice, while the maximum of the pulse decreases with the same
factor.}
\end{figure}

\begin{figure}
\centering
\includegraphics[width=120 mm]{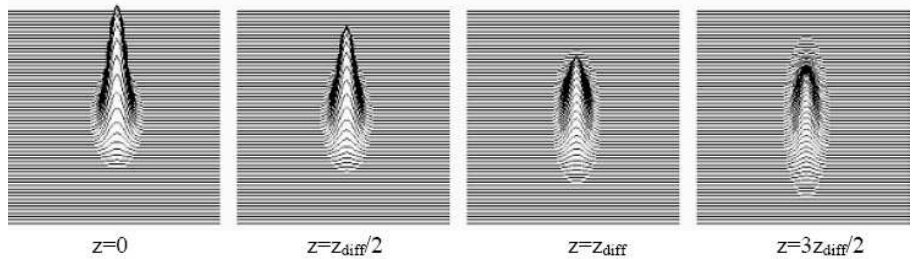}
 \caption{Side ($x,\tau$) projection of the intensity $|A(x,\tau)|^2$ for
the same optical pulse as in Fig. 1. The ($x,y$) projection of the
pulse diffracts considerably following the Fresnel law, while the
($\tau$) projection on several diffraction lengths preserves its
initial shape due to the small dispersion. The diffraction -
dispersion picture, presented by the side $(x,\tau)$ projection,
gives idea of what should happen in the nonlinear regime: the plane
wave diffraction with a combination of parabolic type nonlinear Kerr
focusing always leads to self-focusing for narrow-band ($\triangle
k_z<<k_0$) pulses.}
\end{figure}

\begin{figure}
\centering
\includegraphics[width=120 mm]{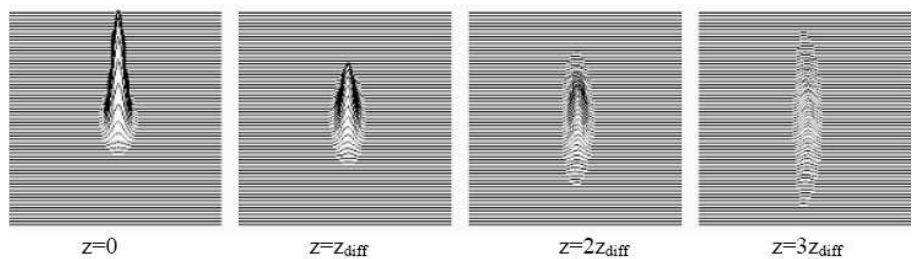}
\caption{Side ($x,z$) projection of the intensity $|A(x,z')|^2$ for
a normalized $10$ fs Gaussian initial pulse at $\lambda=800$ nm,
$\triangle k_z\simeq k_0/3$, $z_0=r_0/2$, and only $3$ cycles under
the envelope (large-band pulse $\triangle k_z \approx k_0$),
obtained numerically from the AE equation (\ref{gal}) in Galilean
frame. At $3$ diffraction lengths a divergent parabolic type
diffraction is observed. In nonlinear regime a possibility appear:
the divergent parabolic type diffraction for large-band pulses  to
be compensated by the converged parabolic type nonlinear Kerr
focusing.}
\end{figure}

b: broad band pulses - from  attosecond up to $20-30$ femtosecond
pulses, where the conditions:

\begin{eqnarray}
\frac{\triangle\omega^2}{v_{gr}^2}\sim k_0^2 \propto k_x^2\sim k_y^2
\end{eqnarray}
are satisfied. In this case we can not use  Taylor expansion of the
spectral kernels in Laboratory (\ref{DDK}), Galilean (\ref{SKB}) and
local time (\ref{locsol}) frames. The spectral kernels are in square
root and  we can expect evolution governed by wave diffraction. That
why for broadband pulses we can expect curvature (parabolic
deformation) of the intensity profile of the  $(x,z)$ or $( x,\tau
)$ side projection. Fig 3.  present the  evolution of the intensity
(side $(x,z)$ projection) of a normalized $10$ fs Gaussian initial
pulse at $\lambda=800$ nm; $\Delta k_z\simeq k_0/3$; $z_0=r_0/2$;
and only $3$ cycles under the envelope (broadband pulse), obtained
numerically from AE equation (\ref{gal}) in Galilean frame. The
solution confirms the experimentally observed parabolic type
diffraction for few cycle pulses. And here appears the main physical
question for stable pulse propagation in nonlinear regime: Is it
possible for the divergent parabolic intensity distribution due to
non-paraxial diffraction  to be compensated by the converged
parabolic type nonlinear Kerr focusing? If this is the case, then a
stable soliton pulse propagation exists. As we show below, only for
broadband pulses one-directional soliton solution of the
corresponding nonlinear equations can be found.

\section{Self-focusing of narrow band femtosecond pulses. Conical emission and spectral broadening}

The laser pulses in a media acquire additional carrier -to envelope
phase (CEP), connected with the group-phase velocity difference. In
air the dispersion is a second order phase effect with respect to
the CEP. In linear regime the envelope equations  contain Galilean
invariance, and thus CEP does not influence  the pulse evolution.
Taking into account the CEP in the expression for the  nonlinear
polarization of third order, a new frequency conversion in THz and
GHz region takes place. In Laboratory frame, the nonlinear
polarization of third order for a laser beam or optical pulse,
without considering CEP, can be written as follows:

\begin{eqnarray}
\label{NLP}
n_2E^3\left(x,y,z,t\right)\vec{x}=\vec{x}n_2\exp{\left[i(k_0(z-v_{ph}t)\right]}\times\nonumber\\
\\
\Biggr\{\frac{3}{4}|A|^2A+
\frac{1}{4}\exp{\left[2i(k_0(z-v_{ph}t)\right]}A^3\Biggr\}+
\vec{x}c.c., \nonumber
\end{eqnarray}
while in Galilean coordinates ($z'=z-v_{gr}t;t'=t$) the CEP, being
an absolute phase \cite{EXTR}, is present in the phase of the Third
Harmonic (TH) term

\begin{eqnarray}
\label{NLPG}
n_2E^3\left(x,y,z,t\right)\vec{x}=\vec{x}n_2
\exp{\left[i\left(k_0(z'-(v_{ph}-v_{gr})t'\right)\right]}\times\nonumber\\
\\
\Biggr\{\frac{3}{4}|A|^2A+
\frac{1}{4}\exp{\left[2i\left(k_0(z'-(v_{ph}-v_{gr})t'\right)\right]}A^3\Biggr\}+
\vec{x}c.c..\nonumber
\end{eqnarray}
Note that  we transform the TH term to a frequency shift of
$\omega_{nl}= 3k_0(v_{ph}-v_{gr})\cong 93 GHz$ in air of the
carrying wave number $\lambda_0=800 nm$. The nonlinear amplitude
equations for power near the critical one for self-focusing in
Laboratory and Galilean frame are:

\begin{eqnarray}
\label{NDE}  -2ik_0\left(\frac{\partial A}{\partial z}+
\frac{1}{v_{gr}}\frac{\partial A}{\partial t}\right) = \Delta A
-\frac{1+\beta}{v_{gr}^2}\frac{\partial^2 A}{\partial t^2}+\nonumber\\
\\
n_2k_0^2 \Biggr\{\frac{3}{4}|A|^2A+
\frac{1}{4}\exp{\left[2i(k_0(z-v_{ph}t)\right]}A^3\Biggr\}+ c.c.,
\nonumber
\end{eqnarray}
and

\begin{eqnarray}
\label{GNDE} -i\frac{2k_0}{v_{gr}}\frac{\partial V}{\partial t'}=
\Delta_{\bot} V-\frac{1+\beta}{v_{gr}^2}\left(\frac{\partial^2
V}{\partial
t'^2}-2v_{gr}\frac{\partial^2 V}{\partial t'\partial z'}\right)+\nonumber\\
\\
n_2k_0^2 \Biggr\{\frac{3}{4}|V|^2V+
\frac{1}{4}\exp{\left[2i\left(k_0(z'-(v_{ph}-v_{gr})t'\right)\right]}V^3\Biggr\}+
c.c. \nonumber,
\end{eqnarray}
respectively. We use AE equations (\ref{NDE}) and (\ref{GNDE}) to
simulate the propagation of a fs pulse, typical for laboratory-scale
experiments: initial power $P=2P_{kr}$, center wavelength
$\lambda=800$ $nm$, initial time duration $t_0=400$ $fs$,
corresponding to spatial pulse duration $z_0=v_{gr}t_0\cong 120$
$\mu m$, and waist $r_0=120$ $\mu m$.
\begin{figure}
\centering
\includegraphics[width=120 mm]{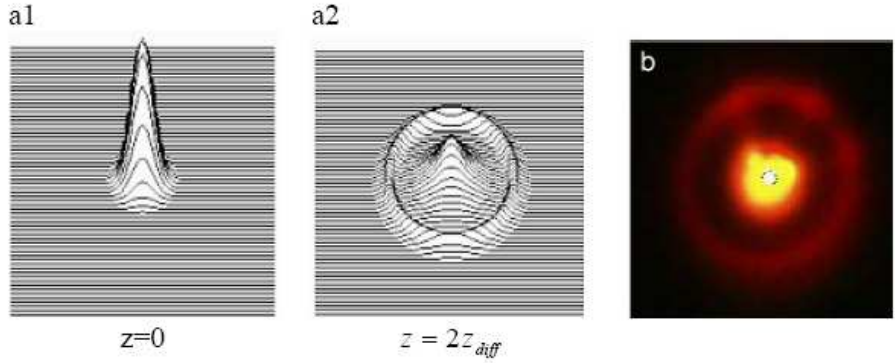}
\caption{Nonlinear evolution of the waist (intensity) projection
$|A(x,y)|^2$ of a $400$ $fs$ initial Gaussian pulse (a1) at
$\lambda=800$ $nm$, with spot $r_0=120$ $\mu m$, and longitudinal
spatial pulse duration $z_0=v_{gr}t_0\cong 120$ $\mu m$ at a
distance $z=2z_{diff}$ (a2), obtained by numerical simulation of the
3D+1 nonlinear AE equation (\ref{NDE}). The power is above the
critical for self-focusing $P=2P_{kr}$ . Typical self-focal zone
(core) surrounded by Newton's ring is obtained. (b) Comparison with
the experimental result presented in \cite{COU}.}
\end{figure}

\begin{figure}
\centering
\includegraphics[width=120 mm]{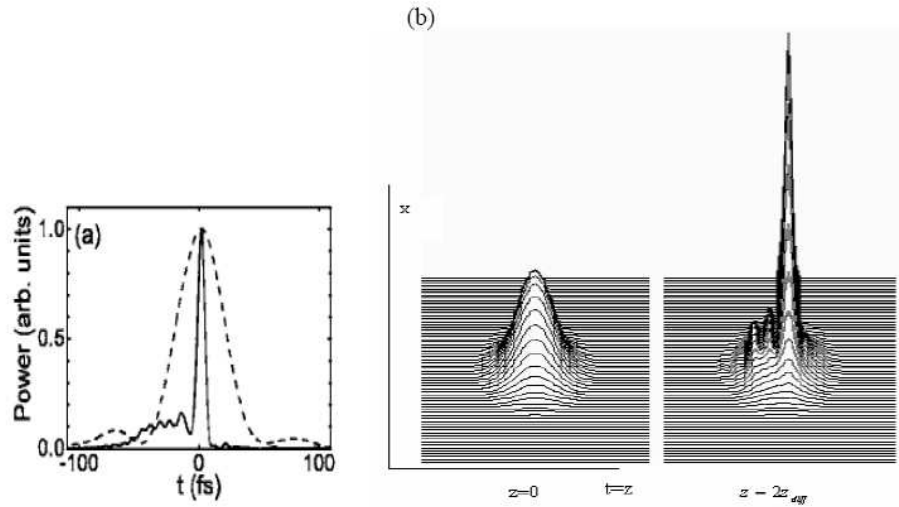}
\caption{(a) Experimental result of pulse self compression and
spliting of the initial pulse to a sequence of several decreasing
maxima  \cite{SKUP}. (b) Numerical simulation of the evolution of
(x,t=z) projection $|A(x,z')|^2 $ of the same pulse of Fig. 4 at
distances $z=0, z=z_{diff}$, governed by the (3D+1) nonlinear AE
equation (\ref{NDE}) and the ionization-free model.}
\end{figure}

\begin{figure}
\centering
\includegraphics[width=120 mm]{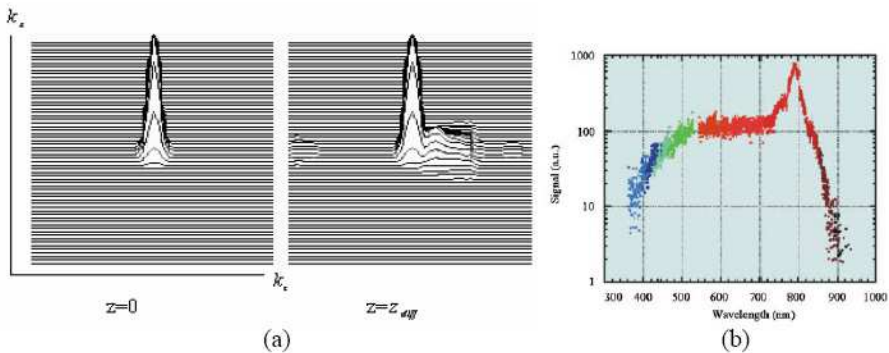}
\caption{(a) Fourier spectrum of the same side ($x,z$) projection of
the intensity $|A(k_x,k_z)|^2$ as in Fig.5. At one diffraction
length the pulse enlarges asymmetrically forwards the short
wavelengths (high $k_z$ wave-numbers),(b) a spectral form observed
also in the experiments  \cite{COU}.}
\end{figure}

Fig.4 presents the evolution of the spot $|A(x,y|^2 $ of  the
initial Gaussian laser pulse at distances $z=0,
z=1/2z_{diff},z=z_{diff},z=3/2z_{diff}$. As a result, we obtain the
typical self-focal zone (core) with colored ring around, observed in
several experiments \cite{COU, KAN, FAC}. The $3D+1$ nonlinear AE
equation (\ref{GNDE}) gives an additional possibility for
investigating the evolution of the side projection of the intensity
$|A(x,z'|^2 $ profile. The side projection $|A(x,z'|^2 $ of the same
pulse is presented in Fig.5. The initial Gaussian pulse begins to
self-compress at about one diffraction length and it is split in a
sequence of  several  maxima with decreasing amplitude. Fig. 6
presents the evolution of the Fourier spectrum of the side
projection $|A(k_x,k_{z'}|^2 $. At one diffraction length the pulse
enlarges asymmetrically towards the short wavelengths (high
wave-numbers).  It is important to point here, that similar
numerical results for narrow band pulses are obtained when only the
self-action term in AE equation (\ref{GNDE}) is taken into account.
The TH or THz term (the second nonlinear term in the brackets)
practically does not influence  the intensity picture during
propagation. In conclusion of this paragraph, we should point out
that our non-paraxial ionization-free model (\ref{NDE}) and
(\ref{GNDE}) is in good agreement with the experiments on spatial
and spectral transformations of a fs pulse in a regime near the
critical $P\geq P_{cr}$. Such transformation of the shape and
spectrum of the fs pulse is typical in the near zone, up to several
diffraction lengths, where the conditions for narrow-band pulse are
satisfied $\triangle k_z<<k_0$.

\section{Carrier-to-envelope phase and nonlinear polarization. Drift from THz to GHz generation}

In nonlinear regime the spectrum of the amplitude function becomes
large due to different nonlinear mechanisms. The Fourier expression
$\hat{A}\left[k_x, k_y,k_0-k_z,\omega_0-\omega\right]$ is a function
of arbitrary $\Delta k_z=k_0-k_z$ and $\Delta
\omega=\omega_0-\omega$, which are related to the group velocity
$\Delta \omega/\Delta k_z=v_{gr}$ (here, we do not include the
nonlinear addition to the group velocity - it is too small for power
near the critical one). Let $\Delta k_z$ denote an arbitrary initial
spectral width of the pulse. In the nonlinear regime $\Delta k_z(z)$
enlarges considerably and approaches values $\Delta k_z(z)\simeq
k_0$. To see the difference between the evolution of narrow-band
$\Delta k_z<<k_0$ and broadband $\Delta k_z\simeq k_0$ pulses, it is
convenient to rewrite the amplitude function in Laboratory
coordinates (the dispersion number  $\beta\simeq 2.1\times10^{-5}$,
being smaller than the diffraction in air, is neglected):

\begin{eqnarray}
\label{b1}
 A\left(x,y,z,t\right)=B_0
{B}\left(x,y,z,t\right)\exp{\left(-i(\triangle
k_z(z-v_{gr}t)\right)},
\end{eqnarray}
while in Galilean coordinates it is equal to:
\begin{eqnarray}
\label{b2}
 V\left(x,y,z',t'\right)=B_0G\left(x,y,z',t'\right)\exp{\left(-i\triangle k_zz'\right)}.
\end{eqnarray}
The Nonlinear Diffraction Equation (NDE) (\ref{NDE}) in Laboratory
frame becomes:

\begin{eqnarray}
\label{NDK}
 -2i(k_0-\triangle k_z)\left(\frac{\partial B}{\partial z}+
\frac{1}{v_{gr}}\frac{\partial B}{\partial t}\right)= \Delta B
-\frac{1}{v_{gr}^2}\frac{\partial^2 B}{\partial t^2}+\nonumber\\
\\
n_2k_0^2B_0^2 \Biggr\{\frac{3}{4}|B|^2B+
\frac{1}{4}\exp{\left[2i\left((k_0-\triangle
k_z)z-(k_0v_{ph}-v_{gr}\triangle k_z)t\right)\right]}B^3\Biggr\}+
c.c. \nonumber,
\end{eqnarray}
and in Galilean frame, the equation (\ref{GNDE}) is:
\begin{eqnarray}
\label{NGDK} -i\frac{2(k_0-\triangle k_z)}{v_{gr}}\frac{\partial
G}{\partial t'}=
\Delta_{\bot}G-\frac{1}{v_{gr}^2}\left(\frac{\partial^2 G}{\partial
t'^2}-2v_{gr}\frac{\partial^2 G}{\partial t'\partial z'}\right)+\nonumber\\
\\
n_2k_0^2B_0^2 \Biggr\{\frac{3}{4}|G|^2G+
\frac{1}{4}\exp{\left[2i\left((k_0-\triangle
k_z)z'-k_0(v_{ph}-v_{gr})t'\right)\right]}G^3\Biggr\}+ c.c.,
\nonumber
\end{eqnarray}
where $\triangle k_z$  can get arbitrary values. It can be seen that
the nonlinear phases in both coordinate systems are equal after the
transformation $z'=z-v_{gr}t;\ t'=t$:

\begin{eqnarray}
(k_0-\triangle k_z)z-(k_0v_{ph}-\triangle k_zv_{gr})t=(k_0-\triangle
k_z)z'-k_0(v_{ph}-v_{gr})t'.
\end{eqnarray}
On the other hand, the corresponding frequency conversions are
different. In Laboratory frame the frequency conversion depends on
the spectral width $\triangle k_z$:

\begin{eqnarray}
\label{WDK} \omega^{Lab}_{nl}=k_0v_{ph}-\triangle k_zv_{gr},
\end{eqnarray}
while in Galilean frame the nonlinear frequency conversion is
fixed to the offset frequency
\begin{eqnarray}
\omega^{Gal}_{nl}=k_0(v_{ph}-v_{gr})= 31 GHz ;\ (\lambda=800nm)
\end{eqnarray}
in air. The expression of the nonlinear frequency shift in
Laboratory frame (\ref{WDK}) explains the different frequency
arising from pulses with different initial spectral width. When the
laser is in ns or ps regime, $\triangle k_z<< k_0$ and the nonlinear
frequency shift is equal to the third harmonic
$3\omega^{Lab}_{nl}=3\omega_0$. In this case (spectral width of the
pulse much smaller than the spectral distance to the third
harmonic), the  phase matching conditions can not be met. Thus, the
nonlinear polarization is transformed into a self-action term. The
fs pulses on the other hand have initial spectral width of the order
$\triangle\omega^{fs}\simeq 10^{13-14} Hz$ and for such pulses at
short distances in nonlinear regime the condition
$\triangle\omega^{fs}\simeq\omega^{Lab}_{nl}$ can be satisfied.
Thus, the nonlinear frequency shift lies within the spectral width
of a fs pulse, and from (\ref{WDK}) follows the condition for THz
and not for TH generation. The self-action enlarges the spectrum up
to values $\triangle k_z\simeq k_0$ and thus, following (\ref{WDK}),
the nonlinear frequency conversion in far field zone drifts from THz
to $\sim 93$ GHz  \cite{MECH}. Note that we consider a single pulse
propagation, while the laser system generates a sequences of fs
pulses. The different pulses have different nonlinear spectral
widths when moving from the source to the far field zone. One would
detect in an experiment a mix of frequencies from THz up to GHz.

\section{Nonlinear sub-cycle regime for $\triangle k_z \approx
k_0$}

The separation of the nonlinear polarization to self-action and
 TH, THz or GHz generated terms is appropriate for fs pulses
up to several cycles under envelope. For fs narrow-band pulses, as
mentioned in the previous section, the pulse shape is changed by the
self-action term, while the CEP frequency depending at the spectral
width of the pulse $\triangle k_z$ leads to different type of
frequency conversion drifts from THz to GHz region. However, when
supper-broad spectrum occurs ($\triangle k_z \approx k_0$), the time
width of the pulse $\triangle t$ becomes smaller than the period of
the nonlinear oscillation $\omega^{Lab}_{nl}$. In this nonlinear
sub-cycle regime, the nonlinear term starts to oscillate with
$\omega^{Lab}_{nl}$ and  separation of the self-action and the
frequency conversion terms becomes mathematically incorrect, due to
the mixing of frequencies \cite{KOL2, KOV2}. For the first time such
possibility was discussed in \cite{KOL2}, where a correct expression
of the nonlinear polarization, including Raman response is
presented. In the sub-cycle regime the nonlinear polarization at a
fixed frequency and Laboratory  frame becomes:

\begin{eqnarray}
\label{NLPL}
n_2E^3\left(x,y,z,t\right)=n_2\exp{\left[i\left((k_0-\triangle
k_z)z-(k_0v_{ph}-\triangle k_zv_{gr})t\right)\right]]}\times\nonumber\\
\\
\Biggr\{\exp{\left[2i\left((k_0-\triangle k_z)z-(k_0v_{ph}-\triangle
k_zv_{gr})t\right)\right]}B^3\Biggr\}, \nonumber
\end{eqnarray}

and in Galilean frame it is

\begin{eqnarray}
\label{NLPG1}
n_2E^3\left(x,y,z',t'\right)=n_2\exp{\left[i\left((k_0-\triangle
k_z)z'-k_0(v_{ph}-v_{gr})t'\right)\right]]}\times\nonumber\\
\\
\Biggr\{\exp{\left[2i\left((k_0-\triangle
k_z)z'-k_0(v_{ph}-v_{gr})t'\right)\right]}B^3\Biggr\}. \nonumber
\end{eqnarray}
In spite of the super-broad spectrum, the dispersion parameter in
the transparency region from $400$ $ nm$ up to $800$ $ nm$ continues
to be small, in the range of $\beta\approx 10^{-4}-10^{-5}$. The
nonlinear amplitude equations for pulses with super-broad spectrum
in Laboratory system become:
\begin{eqnarray}
\label{DKNDE}  -2i(k_0-\triangle k_z)\left(\frac{\partial
A}{\partial z}+ \frac{1}{v_{gr}}\frac{\partial A}{\partial t}\right)
= \Delta A
-\frac{1}{v_{gr}^2}\frac{\partial^2 A}{\partial t^2}+\nonumber\\
\\
n_2k_0^2 \exp{\left[2i\left((k_0-\triangle
k_z)z-(k_0v_{ph}-\triangle k_zv_{gr})t\right)\right]}A^3, \nonumber
\end{eqnarray}
and in Galilean frame

\begin{eqnarray}
\label{DKGNDE} -i\frac{(k_0-\triangle k_z)}{v_{gr}}\frac{\partial
V}{\partial t'}= \Delta_{\bot}
V-\frac{1}{v_{gr}^2}\left(\frac{\partial^2 V}{\partial
t'^2}-2v_{gr}\frac{\partial^2 V}{\partial t'\partial z'}\right)+\nonumber\\
\\
n_2k_0^2\exp{\left[2i\left((k_0-\triangle
k_z)z'-k_0(v_{ph}-v_{gr})t'\right)\right]}V^3  \nonumber.
\end{eqnarray}

\begin{figure}
\centering
\includegraphics[width=100 mm]{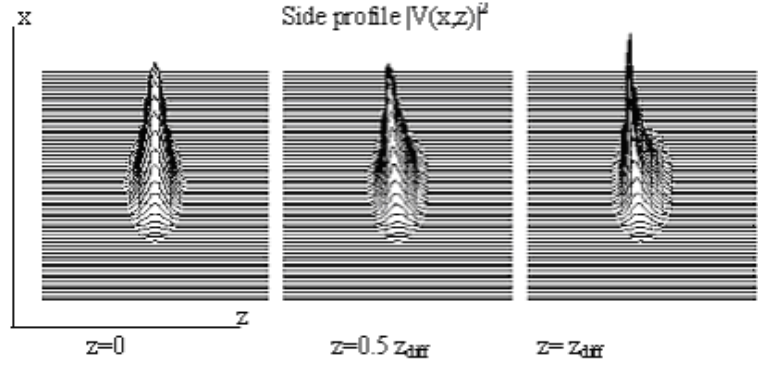}
\caption{Numerical simulations for an initial Gaussian pulse with
super-broad spectrum $\triangle k_z \approx k_0$ governed by the
nonlinear equation (\ref{DKGNDE}). The power is slightly above the
critical $P=2P_{kr}$. The side projection $|V(x,z')|^2 $ of the
intensity is plotted. Instead of splitting into a series  of several
maxima, the pulse transforms its shape into a Lorentzian of the kind
$V(x,y,z')\simeq 1/[1+x^2+y^2+(z'+ia)^2+a^2] $.}
\end{figure}

Fig. 7 shows a typical numerical solution  of the nonparaxial
nonlinear equation (\ref{DKNDE}) (or (\ref{DKGNDE})) for an initial
Gaussian pulse with super-broad spectrum $\triangle k_z \approx
k_0$. It is obtained by using the split step method ($4$ step
Runge-Kutta method for the nonlinear part). These results are the
same both in Laboratory and Galilean coordinate frames differing
only by a translation. The side projection $|V(x,z'|^2$ of the
intensity profile is plotted for different propagation distances.
Instead of splitting into to a series of several maxima, the pulse
transforms its shape in a  Lorentzian type form of the kind
$V(x,y,z)\simeq 1/[1+x^2+y^2+(z'+ia)^2+a^2] $. Here, the number $a$
accounts for compression in $z'$ direction and a spatial angular
distribution. Fig. 8 presents the evolution of the spectrum
$|V(k_x,k_{z'}|^2$ of the  side intensity projection for the same
pulse. The spectrum enlarges forwards the small $k_z$ wave-numbers
(long wavelengths) - typical for Lorentzian type profiles. To
compare with Fig. 8, Fig. 9 gives a plot of the side projection
$|V(k_x,k_z'|^2$ of the spectrum of a Lorentzian profile
$V(x,y,z')=1/[1+x^2+y^2+(z'+ia)^2+a^2]$, $a=2$ increases toward the
small wave-numbers. The numerical experiments lead to the conclusion
that a possible shape of the stable $3D+1$ soliton can be in the
form of a Lorentzian profile. Thus, if we take as an initial
condition  Lorentzian, instead Gaussian one, a relative stability in
the shape and spectrum can be expected. Fig. 10 shows the evolution
of the $|V(x,z'|^2$ profile of a pulse with initial Lorentzian shape
$V(x,y,z',t=0)=1/[1+x^2+y^2+(z'+ia)^2+a^2]$, $a=2$. The pulse
propagates at distance of one diffraction length, preserving its
initial shape.

\begin{figure}
\centering
\includegraphics[width=110 mm]{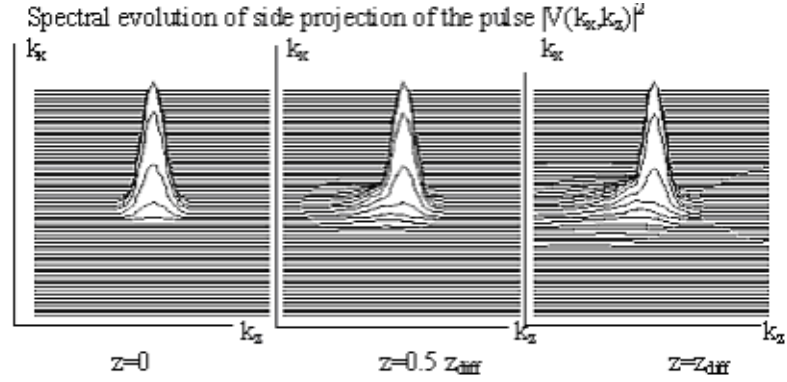}
\caption{The evolution of the spectrum $|V(k_x,k_z)|^2$ of the same
side intensity projection $|V(x,z')|^2 $ . The spectrum enlarges
towards small $k_z$ wave-numbers (long wavelengths) - typical for
Lorentzian profiles.}
\end{figure}

\begin{figure}
\centering
\includegraphics[width=50 mm]{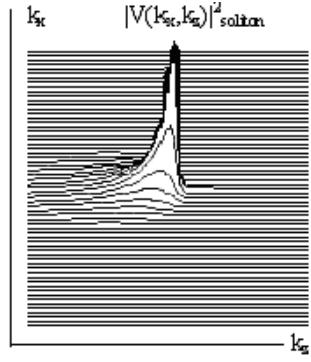}
\caption{Plot of the side projection $|V(k_x,k_z|^2$ of the spectrum
of a Lorentzian profile $V(x,y,z)=1/[1+x^2+y^2+(z+ia)^2+a^2]$, $a=2$
increasing towards the small $k_z$ wave-numbers (compare with Fig.
8).}
\end{figure}

\begin{figure}
\centering
\includegraphics[width=100 mm]{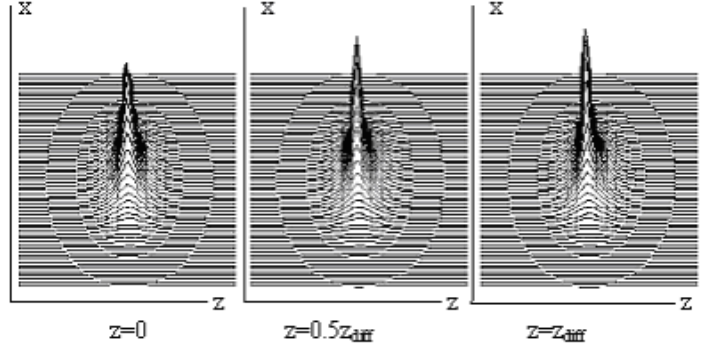}
\caption{Evolution of the $|V(x,z'|^2$ profile of a pulse with
super-broad spectrum $\triangle k_z \approx k_0$ and initial
Lorentzian shape $V(x,y,z',t=0)=1/([1+x^2+y^2+(z'+ia)^2+a^2$],
$a=2$, governed by the nonlinear equation (\ref{DKGNDE}). The pulse
propagates over one diffraction length with relatively stable form.}
\end{figure}

\section{Spectrally  asymmetric 3D+1 soliton solution}

The numerical simulations in the previous section for broad band
spectrum pulses demonstrate a  stable soliton propagation with a
specific initial Lorentzian shape. To find an exact soliton
solution, we require that $\triangle k_z= k_0$ and
$\triangle\omega\cong\omega_0 $  be satisfied. In air $\beta\cong 0$
and the  amplitude equation (\ref{DKNDE}) can be rewritten as:

\begin{eqnarray}
\label{solwave} \Delta B -\frac{1}{v_{gr}^2}\frac{\partial^2
B}{\partial
t^2}+k_0^2n_2B_0^2\exp{\left[i\left(2\triangle\omega_{nl}
t\right)\right]}B^3=0.
\end{eqnarray}
To minimize the influence of the GHz oscillation $\omega_{nl}$, we
use an amplitude function with a phase opposite to CEP:

\begin{eqnarray}
\label{phase} B(x,y,x,t)=C(x,y,z,t)\exp(-i\triangle\omega_{nl} t).
\end{eqnarray}
This corresponds to an oscillation of our soliton solution with
frequency  $\omega_{nl}\simeq 31$ GHz. The equation (\ref{solwave})
becomes:

\begin{eqnarray}
\label{solwave1} \Delta C -\frac{1}{v_{gr}^2}\frac{\partial^2
C}{\partial
t^2}+k_0^2n_2B_0^2C^3=2i\frac{\triangle\omega_{nl}}{v_{gr}^2}\frac{\partial
C}{\partial t}- \frac{\triangle\omega_{nl}^2}{v_{gr}^2}C
\end{eqnarray}
To estimate the influence of the different terms on the propagation
dynamics we rewrite equation (\ref{solwave1}) in dimensionless form.
Substituting:

\begin{eqnarray}
\label{NORM}  t=t_0t;\ z=z_0z;\ x=r_0x;\ y=r_0y;\\
r_0/z_0=\delta\sim 1;\ z_0=v_{gr}t_0;\ t_0\cong2\times
10^{-13}-10^{-14} sec,
\end{eqnarray}
we obtain the following normalized equation:

\begin{eqnarray}
\label{normwave}   \Delta C -\frac{\partial^2 C}{\partial
t^2}+\gamma C^3=i\alpha\frac{\partial C}{\partial t}- \beta C,
\end{eqnarray}
where $\gamma=r_0^2k_0^2n_2B_0^2$ is the nonlinear constant,
$\alpha=2\triangle\omega_{nl}r_0^2/v_{gr}^2t_0$ and
$\beta=\triangle\omega_{nl}^2r_0^2/v_{gr}^2$. For typical  fs laser
pulse at carrier wavenumber $800$ $ nm$ with spot $r_0=100$ $ \mu
m$, the constants of both terms in the r.h.s of equation
(\ref{normwave}) are very small ($\alpha\sim 10^{-2}$ and $\beta\sim
10^{-4}$) and can be neglected. Thus, equation (\ref{normwave})
becomes:

\begin{eqnarray}
\label{Cwave} \Delta C -\frac{\partial^2 C}{\partial t^2}+\gamma
C^3=0.
\end{eqnarray}
Furthermore, we shall assume that the new envelope wave equation
(\ref{Cwave}) has solutions in the form:

\begin{eqnarray}
\label{B} C\left(x,y,z,t\right)=C(\tilde{r}),
\end{eqnarray}
where $\tilde{r}=\sqrt{x^2+y^2+(z+ia)^2-(t+ia)^2}$. From the
nonlinear wave equation (\ref{Cwave}), using (\ref{B}), the
following ordinary nonlinear equation is obtained:

\begin{eqnarray}
\label{rwave}   \frac{3}{\tilde{r}} \frac{\partial
C}{\partial\tilde{r}} +\frac{\partial^2 C}{\partial
\tilde{r}^2}+\gamma C^3=0.
\end{eqnarray}
The number $a$ counts for the longitudinal compression and the phase
modulation of the pulse. When the nonlinear coefficient is slightly
above the critical and reaches the value $\gamma=2$, equation
(\ref{rwave}) has exact particle-like solution of the form:

\begin{eqnarray}
\label{soliton1}   C=\frac{sech(ln(\tilde{r}))}{\tilde{r}}.
\end{eqnarray}
Using the fact that $exp(ln(\tilde{r}))=\tilde{r}$ and
$exp(-(ln\tilde{r}))=\frac{1}{\tilde{r}}$, the solution
(\ref{soliton1}) is simplified to the following algebraic soliton:

\begin{eqnarray}
\label{soliton2} C(\tilde{r})=\frac{2}{1+x^2+y^2+(z+ia)^2-(t+ia)^2}
\end{eqnarray}
The solution (\ref{soliton2}) gives the time evolution of our
Lorentz initial form, investigated in the previous section. As  be
seen from equation (\ref{Cwave}), the solution appears as a balance
between the parabolic (not paraxial) wave type diffraction of broad
band pulse $\triangle k_z= k_0$ and the nonlinearity of third order.
The maxima of this solution are at the points where $\tilde{r}^2=0$.
If we turn back to standard, not normalized coordinates, and solve
the second order equation $z^2+2iaz-2iav_{gr}t-v_{gr}^2t^2=0$, only
one real solution $z=v_{gr}t$ can be obtained. It corresponds to
one-directional propagation with position of the maximum on the $z$
- coordinate $z=v_{gr}t$. As it was pointed  above, Fig. 8 presents
the initial $k_x,k_z$ spectrum of the soliton (\ref{soliton2}).
While the $k_x,k_y$ spectrum is symmetric, the $k_z$ projection is
fully asymmetric, enlarging forwards to low $k_z$ wave-numbers (long
wavelengths), and has typical Lorentz shape. Recently, in
experiments with $2-3$ cycle pulses  long range filaments with
similar spectral profile \cite{KOVACEV} are observed. We suppose
that in this experiment a $3D+1$ Lorentz type soliton was found
experimentally for the first time.

\section{Conclusions}
In this paper we investigate femtosecond pulse propagation in air,
governed by the AE equation, in linear and nonlinear regime. The
equation allows to solve the problem of propagation of pulses with
super-broad spectrum. Note that  this problem can not be studied in
paraxial optics. In linear regime the fundamental solutions of AE
(\ref{DDE}) and DE (\ref{DE}) are obtained and different regimes of
diffraction are analyzed. The typical fs pulses up to $50$ fs
diffract by the Fresnel law, in a plane orthogonal to the direction
of propagation, while their longitudinal shape is preserved in air
or is enlarged a little, due to the dispersion. Broad-band pulses
(only a few cycles under envelope) at several diffraction lengths
diffract in a parabolic form. We solve the convolution problem of
the diffraction equation DE (\ref{DE}) for an initial pulse in the
form of a Gaussian bullet, and obtain an exact analytical solution
(\ref{EXACT1}).  A new method for solving evolution problems of the
wave equation is also suggested. We investigate precisely the
nonlinear third order polarization, including the CEP into account.
This additional phase transforms TH term to THz or GHz terms,
depending on the spectral width of the pulse. Thus, we suggest a new
mechanism of THz and GHz generation from  fs pulses in nonlinear
regime. For pulses with power a little above the critical for
self-focusing, we investigate two basic cases: pulses with
narrow-band spectrum and with broad-band spectrum. The numerical
simulation of the evolution of narrow-band pulses (standard $100$ fs
pulses), gives a typical conical emission and a spectral enlargement
to the short wavelengths. Our study of broad-band pulses leads to
the conclusion that their propagation is governed by the nonlinear
wave equation with third order nonlinear term (\ref{Cwave}), when
the THz oscillation is neglected as small term. An exact soliton
solution of equation (\ref{Cwave}), with $3D+1$ Lorentz shape is
also obtained. The soliton appears as a balance between  parabolic
divergent type diffraction and parabolic convergent type of
nonlinear self-focusing. Numerically, we demonstrate a relative
stability of the soliton pulse with respect to the THz oscillations.

\section{Acknowledgements}

This work is partially supported by the Bulgarian Science
Foundation under grant DO-02-0114/2008.

\section{List of Figure Captions}

Fig.1 Plot of the waist (intensity's) projection $|A(x,y)|^2$ of a
$100$ $fs$ Gaussian pulse at $\lambda=800$ $nm$, with initial spot
$r_0=60$ $\mu m$, and longitudinal spatial pulse duration $z_0=30$
$\mu m$, as solution of the linear equation in local time
(\ref{loct}) on distances expressed by diffraction lengths. The spot
deformation satisfies the Fresnel diffraction law and on one
diffraction length $z=z_{diff}$ the diameter of the spot increases
twice, while the maximum of the pulse decreases with the same
factor.

Fig. 2 Side ($x,\tau$) projection of the intensity $|A(x,\tau)|^2$
for the same optical pulse as in Fig. 1. The ($x,y$) projection of
the pulse diffracts considerably following the Fresnel law, while
the ($\tau$) projection on several diffraction lengths preserves its
initial shape due to the small dispersion. The diffraction -
dispersion picture, presented by the side $(x,\tau)$ projection,
gives idea of what should happen in the nonlinear regime: the plane
wave diffraction with a combination of parabolic type nonlinear Kerr
focusing always leads to self-focusing for narrow-band ($\triangle
k_z<<k_0$) pulses.

Fig. 3 Side ($x,z$) projection of the intensity $|A(x,z')|^2$ for a
normalized $10$ fs Gaussian initial pulse at $\lambda=800$ nm,
$\triangle k_z\simeq k_0/3$, $z_0=r_0/2$, and only $3$ cycles under
the envelope (large-band pulse $\triangle k_z \approx k_0$),
obtained numerically from the AE equation (\ref{gal}) in Galilean
frame. At $3$ diffraction lengths a divergent parabolic type
diffraction is observed. In nonlinear regime a possibility appear:
the divergent parabolic type diffraction for large-band pulses  to
be compensated by the converged parabolic type nonlinear Kerr
focusing.

Fig. 4 Nonlinear evolution of the waist (intensity) projection
$|A(x,y)|^2$ of a $400$ $fs$ initial Gaussian pulse (a1) at
$\lambda=800$ $nm$, with spot $r_0=120$ $\mu m$, and longitudinal
spatial pulse duration $z_0=v_{gr}t_0\cong 120$ $\mu m$ at a
distance $z=2z_{diff}$ (a2), obtained by numerical simulation of the
3D+1 nonlinear AE equation (\ref{NDE}). The power is above the
critical for self-focusing $P=2P_{kr}$ . Typical self-focal zone
(core) surrounded by Newton's ring is obtained. (b) Comparison with
the experimental result presented in \cite{COU}.

Fig. 5 (a) Experimental result of pulse self compression and
spliting of the initial pulse to a sequence of several decreasing
maxima  \cite{SKUP}. (b) Numerical simulation of the evolution of
(x,t=z) projection $|A(x,z')|^2 $ of the same pulse of Fig. 4 at
distances $z=0, z=z_{diff}$, governed by the (3D+1) nonlinear AE
equation (\ref{NDE}) and the ionization-free model.

Fig. 6 (a) Fourier spectrum of the same side ($x,z$) projection of
the intensity $|A(k_x,k_z)|^2$ as in Fig.5. At one diffraction
length the pulse enlarges asymmetrically forwards the short
wavelengths (high $k_z$ wave-numbers),(b) a spectral form observed
also in the experiments  \cite{COU}.

Fig.7 Numerical simulations for an initial Gaussian pulse with
super-broad spectrum $\triangle k_z \approx k_0$ governed by the
nonlinear equation (\ref{DKGNDE}). The power is slightly above the
critical $P=2P_{kr}$. The side projection $|V(x,z')|^2 $ of the
intensity is plotted. Instead of splitting into a series  of several
maxima, the pulse transforms its shape into a Lorentzian of the kind
$V(x,y,z')\simeq 1/[1+x^2+y^2+(z'+ia)^2+a^2] $.

Fig. 8 The evolution of the spectrum $|V(k_x,k_z)|^2$ of the same
side intensity projection $|V(x,z')|^2 $ . The spectrum enlarges
towards small $k_z$ wave-numbers (long wavelengths) - typical for
Lorentzian profiles.

 Fig. 9 Plot of the side projection $|V(k_x,k_z|^2$ of the spectrum
of a Lorentzian profile $V(x,y,z)=1/[1+x^2+y^2+(z+ia)^2+a^2]$, $a=2$
increasing towards the small $k_z$ wave-numbers (compare with Fig.
8).

Fig. 10 Evolution of the $|V(x,z'|^2$ profile of a pulse with
super-broad spectrum $\triangle k_z \approx k_0$ and initial
Lorentzian shape $V(x,y,z',t=0)=1/([1+x^2+y^2+(z'+ia)^2+a^2$],
$a=2$, governed by the nonlinear equation (\ref{DKGNDE}). The pulse
propagates over one diffraction length with relatively stable form.


\begin{thebibliography}{90}

\bibitem{MOU} A. Braun, G. Korn, X. Liu, D. Du, J. Squier, and G.
Mourou, "Self-channeling of high-peak-power femtosecond laser pulses
in air", \ol, 20(1), 73-75 (1995).

\bibitem{WO} L. W\"{o}ste, C. Wedekind, H. Wille, P. Rairoux,
B. Stein, S. Nikolov, C. Werner, S.Nierdermeier, F. Ronneberger, H.
Schillinger, and R. Sauerbrey, "Femtosecond atmospheric lamp",
AT-Fachverlag, Stuttgard, Laser and Optoelectronik 29, 51-53 (1997).

\bibitem{MEC} G. M\'{e}chain, C. D'Amico, Y.-B. Andr\'{e}, S.
Tzortzakis, M. Franco,  B. Prade, A. Mysyrowicz, A. Couairon, E.
Salmon, R. Sauerbrey, "Length of plasma filaments created in air by
a multiterawatt femtosecond laser", \oc, 247, 171-108 (2005).

\bibitem{TZ} S. Tzortzakis, G. M\'{e}chain, G. Patalano, Y.-B. Andr\'{e},
B. Prade, M. Franco, A. Mysyrowicz, J. M. Munier, M. Gheudin, G.
Beaudin, and P. Encrenaz,  "Coherent subterahertz radiation from
 femtosecond infrared filaments in air", \ol, 1944-1946, (2002).

\bibitem{DAMYZ} C. D'Amico, A. Houard, M. Franco,  B. Prade,  A.
Mysyrowicz, "Coherent and incoherent THz radiation emission from
femtosecond filaments in air", Optics Express, 15, 15274-15279
(2007).

\bibitem{DAHO} C. D'Amico, A. Houard, S. Akturk, Y. Liu, J. Le Bloas, M.
Franco,  B. Prade, A. Couairon, V. T. Tikhonchuk, and A. Mysyrowicz,
"Forward THz radiation emission by femtosecond filamentation in
gases: theory and experiment", New J. of Phys., 10, 013015 (2008).

\bibitem{HAU1} C. P. Hauri, W. Kornelis, F. W. Helbing, A. Couairon, A. Mysyrowicz, J.
Biegert, U. Keller, "Generation of intense, carrier-envelope phase
locked few-cycle laser pulses through filamentation", Appl. Phys. B,
79, 673-677 (2004).

\bibitem{HAU2} C. P. Hauri, A. Guandalini, P. Eckle, W. Kornelis, J.
Biegert, U. Keller. "Generation of intense few cycle laser pulses
through filamentation - parameter dependence", Optics Express, 13,
7541 (2005).

\bibitem{COU1} A. Couairon, J. Biegert, C. P. Hauri, W. Kornelis, F. W. Helbing,
U. Keller,  A. Mysyrowicz, "Self-compression of ultrashort laser
pulses down to one optical cycle by filamentation", J. Mod. Opt.,
53, 75-85 (2006).

\bibitem{CHIN1} S.L. Chin, A. Brodeur, S. Petit, O. G. Kosareva, V.
P. Kandidov, " Filamentation and supercontituum generation  during
the propagation of powerful ultrashort laser pulses in optical media
(white light laser)", J. Nonlinear Opt. Phys. Mater., 8, 121-146
(1998).

\bibitem{KASP1} J. Kasparian, R. Sauerbrey, D. Mondelain, S.
Niedermeier, J. Yu, Y. P. Wolf, Y.-B. Andr\'{e}, M. Franco,  B. S.
Prade, S. Tzortzakis, A. Mysyrowicz, H. Wille, M. Rodriguez, L.
W\"{o}ste, "Infrared extenstion of the supercontinuum generated by
femtosecond terrawattlaser pulses propagating in the atmosphere",
\ol, 25, 1397-1399 (2000).

\bibitem{COU} A. Couairon, and A. Mysyrowicz, "Femtosecond filamentation in
transparent media", Physics Reports, 441, 47-189 (2007).

\bibitem{KAN} S. L. Chin, S. A. Hosseini, W. Liu, Q. Luo, F. Th\'{e}berge, N.
Ak\"{o}zbek, A. Becker, V. P. Kandidov, O. G. Kosareva, and H.
Schoeder, "The propagation of powerful femtosecond laser pulses in
optical media: physics, applications, and new challenges", Can. J.
Phys. 83, 863-905 (2005).

\bibitem{FAC} Daniele Faccio, Alessandro Averhi, Antonio Lotti,
Paolo Di Trapani, Arnaud Couairon, Dimitris Papazoglou, Stelios
Tzortzakis, "Ultrashort laser pulse filamentation from spontaneous X
Wave formation in air", Optics Express, 16 1565-1569 (2008)

\bibitem{KOL1} M. Kolesik and J. V. Moloney, "Perturbative and
non-perturbative aspects of optical filamentation in bulk dielectric
media.", Optics Express, 16, 2971-2986 (2008).

\bibitem{SHEN} Y. R. Shen, \textit{The Principles of Nonlinear
Optics}, Wiley-Interscience, New York, 1984.
\bibitem{KASP2} P. B\'{e}jot, J. Kasparian, S. Henin, V. Loriot,
T. Viellard, E. Hertz, O. Faucher, B. Lavorel, and J.-P. Wolf,
"Higher-Order Kerr Terms Allow Ionization-Free Filamentation in
Gases", Phys. Rev. Lett., 104, 103903 (2010).

\bibitem{MECH}  G. M\'{e}chain, A. Couairon, Y.-B. Andr\'{e}, C. D'Amico,
M. Franco,  B. Prade, S. Tzortzakis, A. Mysyrowicz,  R. Sauerbrey,
"Long-range self-channeling of infrared laser pulses in air: a new
regime without ionization", Appl. Phys B, 79, 379-382 (2004).

\bibitem{DUB} A. Dubietis, E. Gai\v{z}auskas, G.
Tamo\v{z}auskas, P. Di Trapani, "Light filaments without
self-channeling", Phys. Rev. Lett 92, 253903 (2004).

\bibitem{BERN} Todd A. Pitts, Ting S. Luk, James K. Gruetzner,
Thomas R. Nelson, Armon McPherson, Stewart M. Cameron and Aaron C.
Bernstein, "Propagation of self-focused laser pulse in atmosphere:
experiment versus numerical simulation", \josab, 21, 2006-2016
(2004).


\bibitem{EXTR} Martin Wegener, \textit{Extreme Nonlinear Optics},
(Springer-Verlag, Berlin Heidelberg, 2005).

\bibitem {HAS} A. Hasegawa, \textit{Optical Solitons in Fibers},
(Springer, Berlin,1989).
\bibitem {AGR} G. P. Agrawal, \textit{Nonlinear Fiber Optics},
(Academic, San Diego, 2001).

\bibitem {SER1} E. M. Dianov, P.V. Mamyshev, A. M. Prokhorov, and V.
N. Serkin, \textit{Nonlinear Effects in Fibers}, (Harwood Academic,
NewYork, 1989).
\bibitem {SER2} V. N. Serkin, A. Hasegawa, and T. L. Belyaeva,
"Nonautonomous Solitons in External Potentials", \prl 98, 074102
(2007).
\bibitem{KOV1} Lubomir M. Kovachev, Kamen  Kovachev,
"Diffraction of femtosecond pulses: nonparaxial regime",\josaa, 25,
2232-2243 (2008);  "Erratum ", 25, 3097-3098 (2008).

\bibitem {CHRIS} I. P. Christov, "Propagation of femtosecond light
pulses", Opt. Comm., 53, 364-366 (1985).

\bibitem {BRAB} T. Brabec, F. Krausz, "Nonlinear Optical Pulse
Propagation in the Single-Cycle Regime", \prl 78, 3282-3285 (1997).

\bibitem{KIS} A. P. Kiselev, "Localized Light Waves: Paraxial and Exact
Solutions of the Wave Equation", Optics and Spectroscopy, 102,
603-622 (2007).

\bibitem{SKUP} Stefan Skupin, Gero Stibenz, Luc Berge, Falk Lederer, Thomas
Sokollik, Matthias Schnurer, Nickolai Zhavoronkov, and Gunter
Steinmeyer ,"Self-compression by femtosecond pulse filamentation:
Experiments versus numerical simulations" Phys. Rev. E 74, 056604
(2006).

\bibitem{KASP3} J. Kasparian, R. Sauerbrey, and S. L. Chin, "The
critical laser intensity of self-guided light filaments in air", \ap
B 71, 877-879 (2000).


\bibitem{KOL2} M. Kolesik, E. M. Wright, A. Becker, and J. V. Moloney,
"Simulation of third-harmonic and supercontinuum generation for
femtosecond pulses in air", \ap B, 85, 531-538 (2006).

\bibitem{KOV2} L. M. Kovachev, "New mechanism for THz oscillation of the
nonlinear refractive index in air: particle-like solutions", J. Mod.
Opt., 56, 1797 - 1803 (2009).

\bibitem{KOVACEV} E. Schulz and M. Kovacev, private communication.
\end{thebibliography}
\end{document}